# Tailoring surface interactions, contact angles, drop topologies and self-assembly using laser irradiation

*John Canning[a\*], Hadrien Weil[b], Masood Naqshbandi[a], Kevin Cook[a] and Matthieu Lancry[b]*

[a] interdisciplinary photonics laboratories (*i*PL), School of Chemistry, The University of Sydney, Sydney NSW 2006, Australia

[b] Institut de Chimie Moléculaire et des Matériaux d'Orsay (ICMMO), UMR CNRS-UPS 8182, Bâtiment 410, Université Paris Sud 11, 91405 Orsay, France

*Corresponding author: E: john.canning@sydney.edu.au  T: 61293511934  F: 61293511911

ABSTRACT UV laser irradiation ($\lambda = 193$ nm), below and above damage thresholds, is used to both alter and pattern the surface properties of borosilicate slides to tune and control the contact angle of a water drop over the surface. Large variation exceeding 25º using laser processing alone, spanning across both sides of the original contact angle of the surface, is reported. An asymmetric contact angle distribution, giving rise to an analogous ellipsoidal-like drop caplet, is shown to improve convective self-assembly of silica nanoparticles into straighter microwires over a spherical caplet.

INTRODUCTION

Omniphilic and omniphobic surface properties play an integral part in determining surface functionality for numerous processes. For microfluidics applications, hydrophobic properties enable water to flow without any significant interactions whilst hydrophilic processes often lead to strong charge based interactions. Wettability to water in particular is of interest to a number of sensing applications where maximum interfacial interaction is important such as biochemical sensing using surface enhanced spectroscopies [1] and many biochemical processes where good attachment is required [2]. Recently, self-assembly of waveguides on hydrophilic glass slides was demonstrated using convective flow within an evaporating drop containing silica nanoparticles [3]. Such convective self-assembly is clearly dependent on the contact angle which must lie in the hydrophilic regime – at the other end of the spectrum, hydrophobic surfaces were used to self-assemble silica nanoparticles into uniform large spheres [4], revealing the topological role of self-assembly and the potential of controlling growth by controlling the shape and volume of the drop. Despite ambiguities as to the nature of the interactions between silica and water, it is reasonable to assume that control of the contact angle offers a degree of control of the self-assembly process. This can be done in various ways with different material systems, whether changing the solvent directly or more practically altering surface properties. For example, magnetic Ni nanowires attached over a surface enabled the demonstration of active tuning between omniphilic and super-omniphobic (contact angle $\alpha > 150°$) behaviour using a magnetic field [5]. Here, we report on the use of laser treatment of glass surfaces (borosilicate and quartz) to demonstrate changes in the wettability (hydrophilic) properties of the glass. Localized laser processing allows patterning of the surface – we therefore demonstrate patterning of the contact angle distribution and use this to control the drop shape from spherical caplet to ellipsoidal-like caplet. The impact on the convective self-assembly of microwires from silica nanoparticles is illustrated. The focus on silica and self-assembling waveguides using bottom up approaches was driven by photonic applications. More importantly, room temperature self-assembly offers a unique possibility of integrating previously incompatible materials, including, for example, organic dyes for sensing [6] and nanodiamonds for possible single photon sources [3].

**THE CONTACT ANGLE AND GLASS-WATER INTERACTIONS**

The wettability of a surface is described by the contact angle, $\alpha$, which is the angle between the solid substrate and the liquid droplet deposited onto it. It is described by Young's equation:

$$\gamma_{SV} - \gamma_{SL} - \gamma_{LV}\cos\alpha = 0 \qquad (1)$$

where $\gamma_{SV}$, $\gamma_{SL}$, $\gamma_{LV}$ are the interfacial tensions between the solid/vapour, solid/liquid and the liquid/vapour states respectively. From the perspective of solid/liquid interactions, this interfacial tension can be affected by changes to the surface properties of the solid; these can be both chemical and physical in nature. Once the solid-liquid interface is affected, in the case of drops on a surface so too will be the liquid/air interface and therefore, we predict, the convective flows during evaporation. Indeed, many different treatments have been employed to change the surface wettability of substrates including: chemical modifications to make superhydrophobic surfaces [7], physical deformation of surfaces by roughening such as the growth of carbon nanotubes on the surface to achieve superhydrophobicity [8] or texturing the surface using nanolithography [9], cleaning methods such as plasma cleaning [10], and laser treatment [11-13]. Even some level of patterning has been reported by controlling microfluidics channels [14].

The interaction between silica and water, which is important for determining whether $\alpha$ will reflect hydrophilic or hydrophobic regimes, is well studied but not altogether clear. Generally, it is widely accepted that the silica surface is already hydroxylated in air, through physisorption, and has hydrophilic properties arising from hydrogen bond interactions between surface hydrogen and water, although this can be readily dehydrated in vacuum [15] or by chemical means [16]. Silica surfaces are often used as a dessicants and make excellent dehydrators [17] so the assumption appears reasonable and is supported by various numerical simulations [18]. The silica/water interaction is nearly always given in terms of removing the hydrogen from a silanol group. Such deprotonation is assumed to occur via hydrogen bonding (SiOH + $H_2O$ + Si→ SiOH···OHH···Si → $SiO^-$ + $H_2O$ + SiH), the principle mechanism by which it is understood leads to a negative surface charge [19]. There is some difficulty in understanding the rationale where the hydride SiH is formed given that the hydroxyl on silica is calculated to be noticeably more stable than the hydride [20-22]. Theoretical calculation using gaseous silicon oxide analogs, for example, estimates the hydroxyl to be up to 200 kJ.mol$^{-1}$ more stable than the hydride [22]. In contrast, FTIR studies on pure silicon reacting with water where both SiH and SiOH form at room temperature and pressure show SiOH, upon annealing >230 ºC, decomposes to form glassy siloxanes (SiOSi) and more SiH [23,24] - this can be explained by noting that the oxidation of pure silicon occurs at much lower energies than either SiOH and SiH formation. At higher annealing temperatures, >300 ºC, 2SiH decompose to release $H_2$. Perhaps on silica it is $H_2$ that is formed with deprotonation: 2(SiOH +$H_2O$) → SiOH···OHH···HHO···OHSi → $2SiO^-$ + $2H_2O$ + $H_2$. It is clear silicon and silica are distinct.

In the case of water drops, an interesting factor which has not been considered is the charge at the water interface and the role of charge balance in determining what happens at the subsequent silica/water interface. Debate, for example, remains as to whether this is positive or negative [25-27] or if it's simply something that arises from restricted orientation of water molecules and therefore dipole alignment at the interface compared to volume. It would appear experimental data supports the argument of $OH^-$ migration through long range force gradients established by the restricted motion at the interface [25], whilst limited numerical simulations restrict calculations to a microscopic level that would support the former, ignoring all possible long range effects. Certainly, the presence of surface $OH^-$ ions would enhance deprotonation by preferentially neutralizing the charge on the water side of the $SiO_2/H_2O$ interface, accounting for the released hydrogen: $SiOH + OH^- \rightarrow SiO^- + H_2O$. In this case, no SiH is necessary. Given the non-uniform chemical terrain of a silica surface, however, it is unlikely that only a single process occurs.

In recent work, we compared different silicate surface modification techniques using the contact angle as a measure of changes at the solid/liquid interface [13]. An untreated pathology grade borosilicate surface produced consistent contact angles $\alpha \sim 27°$, which is clearly hydrophilic. Chemical treatment with Piranha (30% $H_2O_2$: 28% $NH_3$ 28%: 42% $H_2O$) to ensure much greater silanation of the surface increased hydrophilicity, reducing the contact angle to $\alpha \sim 8°$. These experimental data appear consistent with the common assumptions about silica/water interactions. We also found that laser treatment reduces, within experimental error, the contact angle to similar levels, $\alpha \sim 6°$ and, unlike Piranha treatment, is relatively stable over time. The exact reason for this is unclear but laser treatment is known to generate numerous defect centres as well as potentially dehydroxylate a silanated silica surface [28]. A decreased contact angle would suggest that the surface is not dehydroxylated, rather the opposite, if a simple model based on hydrogen bonding as the key source of water attraction is employed. Alternatively, this result indicates that surface silanation with Piranha is not straightforward, and some deprotonation may be occurring with moisture in the air. The role of cleaning the surface of residual species, whilst it cannot be entirely ruled out, was dismissed as being insignificant after obtaining no change, within experimental error, between a sample sonicated in ethanol and the untreated sample. Laser processing at 193nm will also increase defect sites, generating various oxygen deficiency centres – nitrogen gas was used during the process to keep the sample dry, presumably reducing the likelihood of additional silanation from the air.

Given the stability of this process afterwards, we suspect that there is more than one process involved with determining the overall hydrophilic and hydrophobic properties of glass. In addition to various chemical pathways, the effects of double layer interactions, dipole alignment and charge balancing in the liquid phase cannot be discounted. To demonstrate how complex this process likely is, we loaded our silicate samples with molecular hydrogen (80 °C, 180 atm) and observed a significant increase in contact angle from the pristine case to $\alpha \sim 50º$ [13]; i.e. hydrogen generally would lead to increased hydrophobicity, in contrast to hydroxylation.

In this work, we focus on using laser processing to alter the contact angle distribution around a drop; in other words we propose and demonstrate that laser processing can fine tune the interfacial solid-liquid interface around a drop to alter its effective geometry. In the small drop domain limit, we show how the spherical caplet formed on a hydrophilic surface can be made ellipsoidal and the impact this has on convective self-assembly of silica nanoparticles within the drop.

**RESULTS AND DISCUSSION**

Samples were prepared in accordance with our previous work: (1) pristine sample with no surface treatment; (2) samples which are laser treated (details in Materials and Method). To perform the contact angle measurements, 10 µl droplets of $H_2O$ were deposited onto the surface of all samples. The results are summarised in Figure 1 – despite different levels of glass impurities, all measurements are consistent with that reported in [13] but with slightly lower pristine contact angles, $\alpha \sim (24 \pm 5)°$, and slightly higher laser treated contact angles: $\alpha \sim (8 \pm 1)°$, $(10 \pm 1)°$. The observed reduction in contact angle is important – it again indicates for the energy regime used in this work that the silica has not been dehydroxylated since the surface is more hydrophilic, in contrast to that observed for lower energy 255 nm light from a copper vapour laser [28]. However, the relatively long lifetime of the process is consistent with the observation of a long term change through laser processing. A small degradation in the laser processed surface was observed 17 hours later; it may be explained by either moisture from the air or trace amounts of dust. This provides confidence in the measurement setup overall and reproducibility as well suggesting that the properties are generic. Of all the treatments, only laser processing allows straightforward patterning of the surface and it is clearly the most stable over time so it is this process that we focus on.

*Contact angle and laser intensity*

To assess the energy dependence, the contact angle was measured for both borosilicate and quartz samples treated with various energies. Quartz has no dopants and a higher damage threshold; despite this it has shown similar results to borosilicate; interestingly, the contact angles are noticeably lower for quartz pristine case at $\alpha \sim (20 \pm 1)°$, suggesting greater silanation if that was the only mechanism involved (the role of defects is clearly ambiguous given there are likely to be less). The laser processing conditions are identical to that as above (described in detail in Materials and Methods) with only the energy being varied from $E = 0$ *to* $E = 250$ mJ/cm$^2$ focused into a line area which is subsequently scanned over 1cm$^2$. Figure 2 shows the contact angle as a function of laser energy in (mJ/cm$^2$) for pure quartz and borosilicate glass slides. Below $E < 100$ mJ/cm$^2$ for borosilicate and $E < 150$ mJ/cm$^2$ for quartz, there is no significant change in $\alpha$. Above this "threshold", which is not characterized by significant damage other than some coloring in the borosilicate samples (no substantial surface roughening), the change is rapid. It is observed that borosilicate glass slide has a lower "threshold" energy ($E_{th} \sim$ 125 mJ/cm$^2$), where a rapid reduction of $\alpha$ appears, well before that of quartz ($E_{th} \sim 175$ mJ/cm$^2$). This reduction is significant ($\Delta\alpha \sim -10°$ for borosilicate and $\Delta\alpha \sim -6°$ for quartz), in both cases making the surface more hydrophilic. Although $\alpha$ steadily decreases with increasing energy, it is at the threshold, or onset, energy that a significant drop is observed. An alternative explanation is that this change at $E_{th}$ could be the energy required to blast any residual species present at the surface; i.e. laser cleaning. By removing these properly, the wettability is increased. Within experimental error, once the threshold is exceeded there is a similar rate of reduction in contact angle, *dα/dE*, between borosilicate and quartz suggesting a common silica-based mechanism.

*Laser patterning*

To demonstrate and observe the effects of non-uniform treatment of the surface, parallel lines were exposed under various irradiation conditions in an attempt to find an initial optimum contact angle variation, $\Delta\alpha$, between $\alpha=$ measured parallel to the laser lines and $\alpha||$ measured orthogonal to the lines ($\Delta\alpha = \alpha|| - \alpha=$). In contrast to the exposed areas, no scanning occurred – this led to surface interference effects generating a textured layer up to a peak energy beyond which ablation leads to deeper tracks. Whereas in the previous case, the same contact angle is measured throughout the sample irradiated over a 1 cm$^2$ region, on these samples a clearly discernible difference in contact angle, depending along which plane it is measured, is anticipated. The results for $α||$ of three line separations (0.1, 0.5 and 1 mm) on the borosilicate slides are summarized in Figure 3. In all cases, the observation is of an increase in contact angle, $α||$, the exact opposite observed for a continuous bulk exposure – in other words, although the UV exposure itself increases the local hydrophilicity of the surface by reducing the contact angle, the textured surface and spacing generates a periodic variation in contact angle which plays a more important role in affecting the drop interaction with the surface. The peak contact angle corresponds to the point after which the textured patterning is lost with ablation leading to deep damage tracks. Figure 4 illustrates scanning electron micrographs below and beyond an effective ablation threshold. The physical phenomena for such textured surfaces can be explained by plasma generation and exciton interactions that interfere to generate sub-micron or nanoscale surface ripple effects [29, 30]. Directionality of the surface texture could be obtained by controlling the 193nm laser polarisation. Interestingly, the parallel measurements show no change in contact angle indicating the texture contribution is negligible in this direction. The interfacial tension must be affected strongly – for an optimum spacing ~ 0.5 mm, the contact angle increases to 35º, compared to 23º of the untreated sample and 10º for the bulk laser exposure. This is as much as 25º between uniform and patterned processing, offering a powerful tool for surface shaping.

For the borosilicate glass slide, the contact angle as a function of energy when the line spacing is 0.5 mm is shown in Figure 5. At low energies below any damage (<150 mJ/cm$^2$), it was found that the parallel contact angle was in fact constant, $α$= ~ (23 ± 2)º, whilst the contact angle measured orthogonal to the laser line increases as energy is increased from $α||$ ~ 23º to 28º. Thus, for 0.5 mm spacing at energies $E < 150$ mJ/cm$^2$, the change in contact angle around the drop is $\Delta\alpha \sim (0-5)°$. The previously spherical caplet is now ellipsoidal, although this was visually difficult to ascertain from the shape alone. In contrast to the results obtained for uniform exposure area, there is an increase in contact angle overall, which indicates that the patterned surface distribution has increased hydrophobicity. This is suggestive of a sensitive, yet relatively coarse, topological role dominating over any microscopic mechanism.

At much higher energies, larger changes are observed: a peak is obtained when $E = 342$ mJ/cm$^2$, where $\Delta\alpha = (11.6 \pm 1.3)°$. This is consistent with increasing damage in the sample (see the inset images of Figure 5) – at higher energies ablation leads to deep tracks whilst at lower energies the textured surfaces increases hydrophobicity. This transition is confirmed by SEM images shown in Figure 4 for two energies on either side of the peak in Figure 5. Interestingly, despite overt browning of the glass overall from stray low intensity 193nm light, it is observed that the contact angle measured parallel to the laser line remains constant, $\alpha = \sim (23 \pm 2)°$, whilst the contact angle measured orthogonal to the laser line increases to as high as $\alpha_{||} \sim 35°$ before again decreasing to a constant level of $\alpha_{||} \sim 28°$. The spacing in-between the lines was as important if note more than the texturing of the surface within the lines in optimising the increase in hydrophobic contribution.

*Convective self-assembly in within an ellipsoidal caplet droplet*

The ability to control the contact angle of surfaces using laser treatment, both to make them more or less hydrophilic (or hydrophobic) by laser treatment either with continuous exposures or various patterns in two dimensions (e.g. lines) opens up new opportunities for controlling a variety of processes. In this section, an example of how it can be utilized to improve self-assembly is provided. A simpler configuration to demonstrate the potential of this method is applied to the convective self-assembly of silica microwires [3]. The drop is deposited over a rectangular exposed area where it overlaps into the untreated regions as illustrated in Figure 6(b) for the case when $E = 466$ mJ/cm$^2$. The reason for doing this is as follows: from our previous work, the more the contact angle can be reduced the better the wire formation; i.e. convective self-assembly of uniform wires prefers a more hydrophilic surface. We therefore choose to reduce the contact angle and the above experiments indicate this is done by using uniform exposures rather than lines. The exposure conditions of this area are selected on the basis of the previous experiments: one experiment is undertaken at lower energies $E \sim 173$ mJ/cm$^2$ and the other at higher energies $E \sim 466$ mJ/cm$^2$ where damage is obvious. The area of exposure is $A \sim$ 1cm$^2$ as previously and the drops are deposited on one side. Drops of aqueous silica nanoparticle solution were used ($\phi \sim$ 20-30 nm measured by dynamic light scattering and SEM [3], 50 μL, 5% w/w, trace NH$_4^+$; without NH$_4^+$ aggregation of particles is observed to precipitate in solution suggesting no role for coulombic repulsion of charged water/silica, or dipole, interfaces). These were deposited on both a pristine and a laser-irradiated glass slide using an automated pipette and left to dry at ambient conditions ($T = 21°$ C, $P = 1$ atm), as shown in Figure 6. Before drying the contact angles were measured at two energies, summarised in Figure 7. Notably for both cases the contact angle between the water droplet and the laser processed (LP) region is lower ($\alpha = 22 \pm 2°$) than that measured on the unexposed (U) region ($\alpha = 17 \pm 2°$). This supports our previous results showing that laser treatment of a uniform area tends to decrease the contact angle, i.e. make the surface more hydrophilic. The results also further confirm the ability to arbitrarily change the contact angle around a drop.

During evaporation, micro wires and tapers self-assemble – the process is driven by growing radial stresses at the evaporating drop front and resistant van der Waals forces which leads to fracture planes. The shape and uniformity of these fracture planes is affected by drop size and in the spherical caplet regime they are usually tapered [3,6]. After evaporation these are examined under a microscope and dimensions taken – the results are shown for the 466 mJ/cm$^2$ case in Figure 8; qualitatively similar results were obtained at the lower energy. For the sample prepared on the untreated surface, a uniform approximately spherical caplet is formed (figure 8(a)) and, for relatively small drop sizes, this leads to tapered wires with little uniformity pointing radially inwards to the centre of where the drop had been (Figure 8(a)). The crater at the centre is a clear sign of circular convective flow within the drop during evaporation and is a very good measure of the spherical topology of the drop and process. However, in the case where the droplet lies on the laser treated area (Figure 8(b)), after evaporation a clear ellipsoidal crater (figure 8(b)) is strong evidence for asymmetric convective flow caused by the perturbed topology generated by the altered contact angle at the laser exposed region. This leads to much more uniform wire formation as the radial stresses are now focused at two points and within the region in-between there is an equilibration of forces for most of the self-assembly growth. In this case, growth is most uniform towards, or from, the laser processed region. Much more uniform wires are observed and measured as is shown comparing between Figures 8(a1),(a2) and 8(b1),(b2).

**CONCLUSIONS**

Laser processing has been demonstrated to be effective for both fine tuning and patterning surface properties, opening an extraordinary new tool for shaping surface properties and directing processes that depend on surface interactions. The degree of hydrophilic contact angle of a water drop on silica was varied significantly over a range from 6º to as high as 35º; bulk processing by scanning tends to increase the hydrophilicity of the surface whereas patterning using fixed exposure lines can decrease this (or increase hydrophobicity). Where scanning is involved, the texture pattern is effectively washed out. For the large area exposures, despite surface damage by the laser itself, within the energies and laser parameters used here we did not observe any increase in hydrophobicity; this was only observed with patterning of the surface using lines. This difference is explained in part by surface texturing arising from laser and plasma/exciton interference up to a peak in hydrophobic increase beyond which ablation occurs and a decline is observed. It is also enhanced by optimising the spacing between lines. Such patterning generates asymmetric contact angle distributions – a combination of the two may offer further scope for greater control and finesse. Under such conditions, surface area may also be increased by generating porous zeosil glass at the same time [31].

Whilst at first glance laser patterning induced asymmetry appears visually small in terms of the conversion of a spherical caplet to an approximate ellipsoid caplet, the corresponding distribution in solid/liquid tensions is sufficient enough to affect convective flow patterns during evaporation. This was demonstrated by showing how self-assembly of microwires within the ellipsoidal caplet regime creates significantly more uniform wires under such conditions – it is an example of how a process has also been made directional. We can therefore conclude that the topological shape of the drop is an indicator of different interface tensions, including at the liquid/air interface, and that this is by far the most important process for controlling convective flow within a liquid drop. Indeed, by tailoring these processes, a sophisticated tool now exists to control flow within drops and other microfluidics formats of value to a range of technologies from sensing, micro(nano)-mixing and micro(nano)fluidics, to convective self-assembly of components such as the photonic microwires demonstrated here. Likewise, directional chemical interactions distributed on a surface can be explored.

The results generally also indicate a greater need to better understand the dynamic interaction between the contact angle and physical phenomena. It is clear that the common assumption between contact angle and normal hydrophilic material properties is more complicated in the presence of patterns and non-uniform exposures and a degree of caution linking what is essentially a complex parameter to such intrinsic properties is required. Nevertheless, this complexity opens up new opportunities for novel and advanced surface science and fabrication.

**MATERIAL AND METHODS**

The substrates used in these experiments are all standard, low-cost microscope borosilicate glass slides (Sail Brand; typical composition: 80% $SiO_2$, 14% $B_2O_3$, 4% $Na_2O$, 2% $Al_2O_3$) and for comparison quartz slides.

The contact angles, averaged over ten samples to obtain a standard deviation, were measured using a home-made contact angle goniometer. A CMOS camera and lens, integrated within a commercial portable field microscope, were used to image these drops from the side (Figure 9). Digital software analysis of the images on a portable netbook was used to determine the contact angle, enabling the entire arrangement to be field deployable. To ensure reproducibility and a reasonable standard deviation, the measurements were performed 10 times each. The equipment was suitable for measuring contact angles greater than 2º with the best standard deviation reducing to ~ ±1º at low contact angles.

Guided by our previous work [13], a similar routine, where the contact angle was measured ten times, was followed between pristine and laser treatments on the borosilicate slides:

(1)   Pristine slide - no surface treatment;

(2)   A slide was laser treated ($\lambda$ = 193 nm, $E_{pulse}$ = 160 mJ/cm$^2$, beam A = 0.0125 cm$^2$, $RR$ = 100Hz, scan $v$ = 1mm/min), over an area $A$ = 1 cm$^2$ by scanning the line profile. For the case of line exposures no scanning took place preventing surface interference from erasing to produce a patterned profile at some energies and a damage track at higher energies. N$_2$ was blown over the surface during irradiation to minimise any oxidative effects and reduce silanation after dehydroxylation. These energies were selected to be below any significant damage threshold to ensure there is no two-photon excitation for the band edge which can lead to surface topology changes that will only help to confuse the interpretation of results. Some photochromic changes are observed as light coloring of the borosilicate slides.

**Author Contributions**

The manuscript was written through contributions of all authors. All authors have given approval to the final version of the manuscript.

**Funding Sources**

The work reported here was funded by Australian Research Council (ARC) grant: FT110100116. Hadrien Weil acknowledges the Agence Nationale pour la Recherche (ANR-09-BLAN-0172-01) and Marie Curie FP7-PEOPLE-IRSES e-FLAG for supporting and funding his visit to Australia. M. N would like to thank The University of Sydney for a postgraduate Gritton Scholarship.

*The authors declare there is no financial or other conflict of interest involved with this work.*

REFERENCES

1. Ataka, K.; Heberle, J. Biochemical applications of surface-enhanced infrared absorption spectroscopy. *Anal. Bioanal. Chem*. **2007,** *388*, 47-54

2. Liu, X.; Chu, P. K.; Ding, C. Surface modification of titanium, titanium alloys, and related materials for biomediacl applications. *Mat. Sci. Eng.* **2004**, *47*, 49–121

3. Naqshbandi, M.; Canning, J.; Gibson, B. C.; Crossley, M. J. Room temperature self-assembly of mixed nanoparticles into complex material systems and devices. *ArXiv* **2012** arXiv:1201.3815v1 [cond-mat.mtrl-sci]

4. Naqshbandi, M.; Canning, J.; Lau, A.; Crossley, M. J. Controlled Fabrication of Macroscopic Mesostructured Silica Spheres for Potential Diagnostics and Sensing Applications. *The Int. Quantum Electronics Conf. (IQEC)/Conf. on Lasers & Electro-Optics (CLEO) Pacific Rim, (IQEC/CLEO-Pacific Rim 2011)*, 2011, Sydney, Australia.

5. Grigoryev, A.; Tokarev, I.; Kornev, K. G.; Luzinov, I.; Minko, S. Superomniphobic Magnetic Microtextures with Remote Wetting Control. *J. Am. Chem. Soc.* **2012** *134*, 12916-12919.

6. Naqshbandi, M.; Canning, J.; Crossley, M. J. Self-assembled silica microwire: a new platform for optical sensing. *OSA Congress, Optical Sensors*, Monterey, California, USA 2012, paper Stu4F.

7. Larmour, I. A.; Bell, S. E. J.; Saunders, G. C. Remarkably simple fabrication of superhydrophobic surfaces using electroless galvanic deposition. *Angew. Chemie*, **2007,** *119*, 1740–1742.

8. Lau, K. K. S.; Bico, J.; Teo, K. B. K.; Chhowalla, M.; Amaratunga, G. A. J.; Milne, W. I.; McKinley, G. H.; Gleason, K. K. Superhydrophobic carbon nanotube forests. *Nano Lett.* **2003,** *3*, 1701–1705.

9. Park, K-C.; Choi, H. J.; Chang, C-H.; Cohen, R. E.; McKinley, G. H.; Barbatathis, G. Nanotextured silica surfaces with robust superhydrophobicity and omnidirectional broadband supertransmissivity. *NANO* **2012** *6*, 3789-3799.

10. Amin, N.; Cheah, A. Y.; Ahmad, I. Effect of plasma cleaning process in the wettability of flip chip PBGA substrate of integrated circuit packages. *J. Appl. Sci.* **2010,** *10*, 772-776.

11. Triantafyllidis, D.; Li, L.; Stott, F. H. The effects of laser-induced modification of surface roughness of $Al_2O_3$-based ceramics on fluid contact angle. *Mat. Sci. Eng. A*, **2005** *390*, 271-277.

12. Sun, C.; Zhao, X. W.; Han, Y. H.; Gu, Z. Z. Control of water droplet motion by alteration of roughness gradient on silicon wafer by laser surface treatment. *Thin SO. Films*, **2008** *516*, 4059-4063.

13. Canning, J.; Petermann, I.; Cook, K. Surface treatment of silicate based glass: base Piranha treatment versus 193nm laser processing. *Proc. SPIE* **2012,** *851*, 83512N.

14. Abate, A. R.; Thiele, J.; Weinhart, M.; Weitz, D. A.; Patterning microfluidic device wettability using flow confinement, *Lab Chip* **2010** *10*, 1774-1776.

15. Chen, Y-W.; Cheng, H-P. Interaction between water and defective silica surfaces. *J. Chem. Phys.* **2011**, *134*, 114703

16. Yoronin, E. F.; Pakhlov, E. M.; Chuiko, A. A. Effect of dehydration of a silica on chemisorptions of methanol. *J. Appl. Spec.* **1997,** *54*, 315-318 (and refs therein).

17. Van Gemert, R. W.; Cuperus, F. P. Newly developed ceramic membranes for dehydration and separation of organic mixtures by pervaporation. *J. Membr. Sci.* **1995**, *105*, 287–291.

18. Fogarty, J. C.; Aktulga, H. M.; Grama, A. Y.; van Dulin, A. C. T.; Pandit, S. A. A reactive dynamics simulation of the silica-water interface. *J. Chem. Phys*. **2010**, *132,* 174704.

19. Behrens, S. H.; Grier, D. G. The charge of glass and silica surfaces. *ArXiv* **2001**, ArXiv:cond-mat/0105149x2 [cond-mat.soft].

20. Bruna, P. J.; Grein, F. MRD-CI study on the isomers SiOH and HSiO 1. Relative stability and electronic spectra. *Molec. Phys.* **1988**, *63,* 329-349.

21. Yamaguchi, Y.; Schaefer, H. F. The $SiOH^+$- $HSiO^+$ system: a high level ab initio quantum mechanical study. *J. Chem. Phys*. **1995**, *102,* 5327-334.

22. Darling, C. L.; Schlegel, H. B. Heats of formation of $SiH_nO$ and $SiH_nO_2$ calculated by ab initio molecular orbital methods at the G-2 level of theory. *J. Phys. Chem*. **1993**, *97,* 8207-8211.

23. Gupta, P.; Dillon, A. C.; Bracker, A. S.; George, S. M. FTIR studies of $H_2O$ and $D_2O$ decomposition on porous silicon surfaces. *Surface Sci.* **1991**, *245*, 360-372.

24. Flowers, M. C.; Jonathan, N. B. H.; Morris, A.; Wright, S. The adsorption and reactions of water on Si(100)-2 x 1 and Si(111)-7 x7 surfaces. *Surface Sci.* **1996**, *351*, 87-102.

25. Gray-Weale, A.; Beattie, J. K. An explanation for the charge on water's surface. *Phys. Chem, Chem. Phys*. **2009**, *11,* 10994011005.

26. Vacha, R.; Horinek, D.; Buchner, R.; Winter, B.; Jungwirth, P. Comment on "An explanation for the charge on water's surface", by A. Gra-Weale and J.K. Beattie, Phys. Chem. Chem. Phys., 1009, 11, 10994. *Phys. Chem, Chem. Phys.* **2010**, *12*, 14362-14363.

27. Gray-Weale, A.; Beattie, J. K. Reply to the 'Comment on "An explanation for the charge on water's surface"' by R. Vacha, D. Horinek, R. Buchner, B. Winter, P. Jungwirth, Phys. Chem, Chem. Phys., 12, 14362-14363 (2010). *Phys. Chem, Chem. Phys.* **2010**, *12*, 14364-14366.

28. Halfpenny, D. R.; Kane, D. M.; Lamb, R. N.; Gong, B. Surface modification of silica with ultraviolet laser radiation. *Appl. Phys*. **2000**, *A71,* 147-151.

29. Birnbaum, M. Semiconductor surface damage produced by ruby lasers. *J. Appl. Phys*. **1965**, *36*, 3688–3689.

30. Beresna, M.; Gecevičius, M.; Kazansky, P. G.; Taylor, T.; Kavokin, A. V. Exciton mediated self-organisation in glass driven by ultrashort light pulses. *Appl. Phys.* **2012**, *101,* 053120.

31. Canning, J.; Lancry, M.; Cook, K.; Poumellec, B. Zeosil formation by femtosecond laser irradiation. *Bragg Gratings, Photosensitivity, and Poling in Glass Waveguides, OSA Technical Digest* (online) (Optical Society of America, 2012) paper BW1D.5.

BRIEFS (Word Style "BH_Briefs"). If you are submitting your paper to a journal that requires a brief, provide a one-sentence synopsis for inclusion in the Table of Contents.

SYNOPSIS (Word Style "SN_Synopsis_TOC"). If you are submitting your paper to a journal that requires a synopsis, see the journal's Instructions for Authors for details.

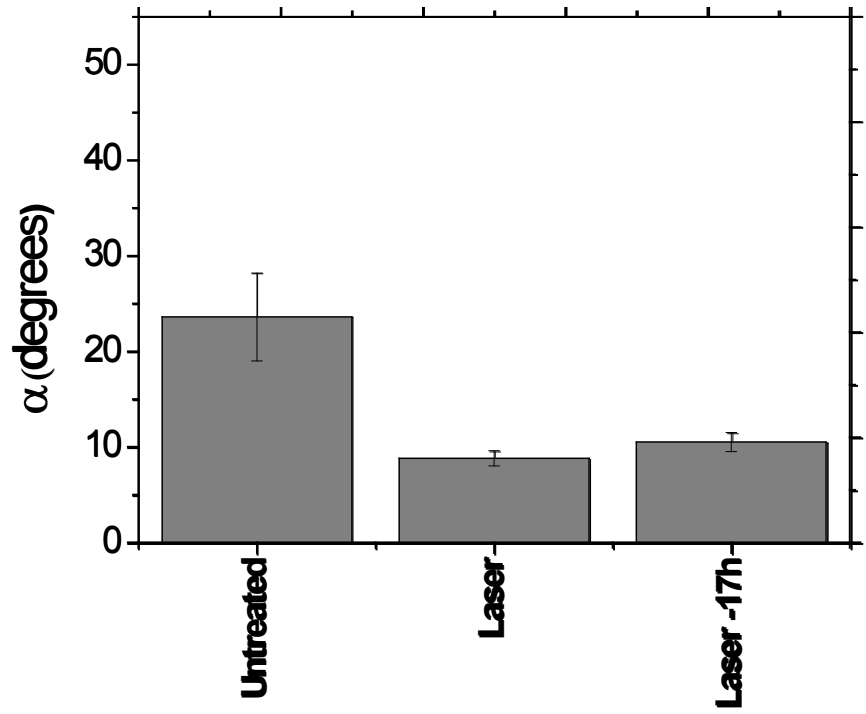

**Figure 1.** Comparison of contact angle $\alpha$ between untreated and laser processed borosilicate.

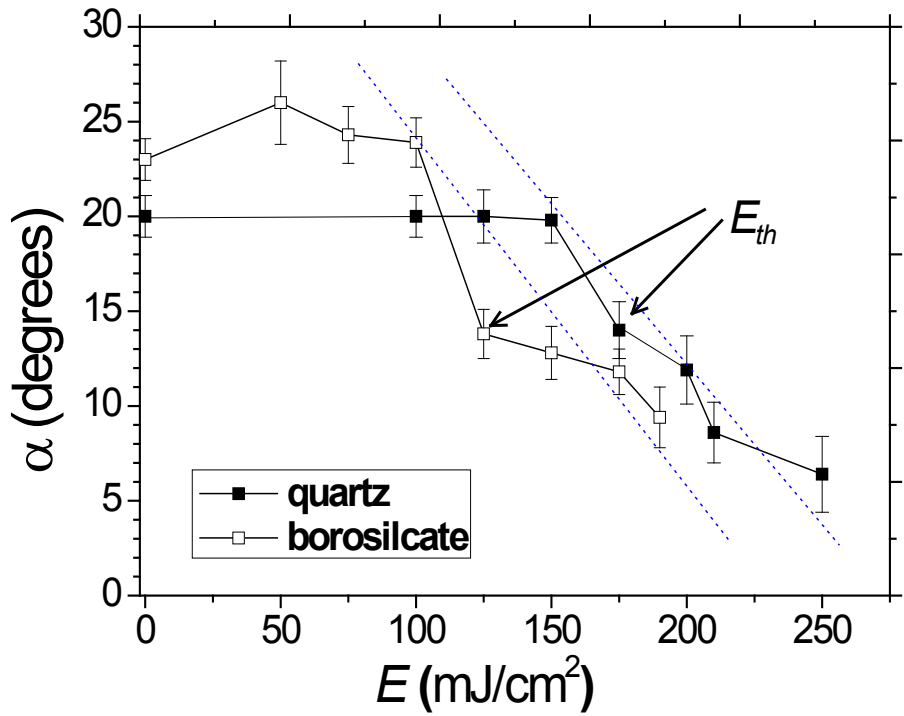

**Figure 2.** Comparison in contact angle $\alpha$ between quartz and borosilicate as a function of laser beam energy. Lines of best fit are shown for the region where change occurs.

.

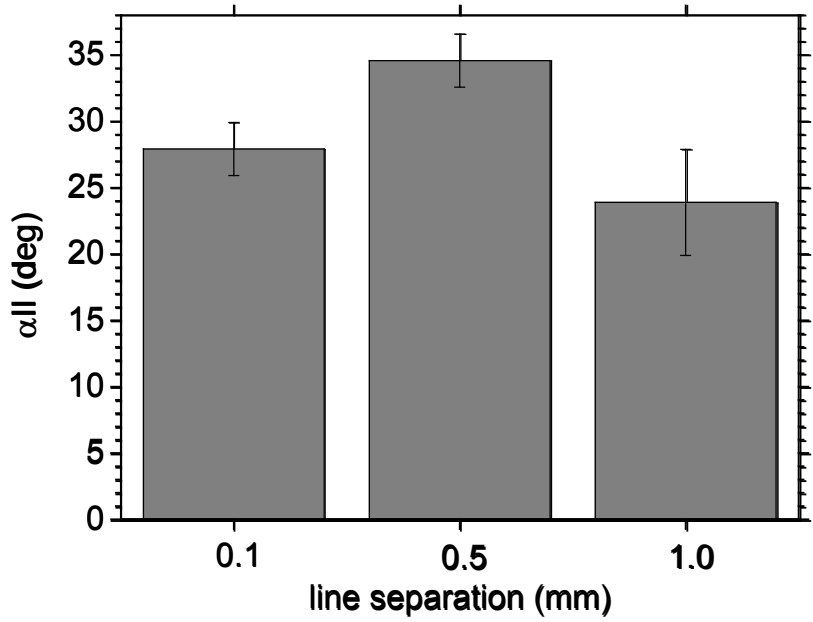

**Figure 3.** Contact angle for three different spacing of 193nm laser processed lines on a borosilicate glass slide. A maximum value is obtained closer to 0.5 mm.

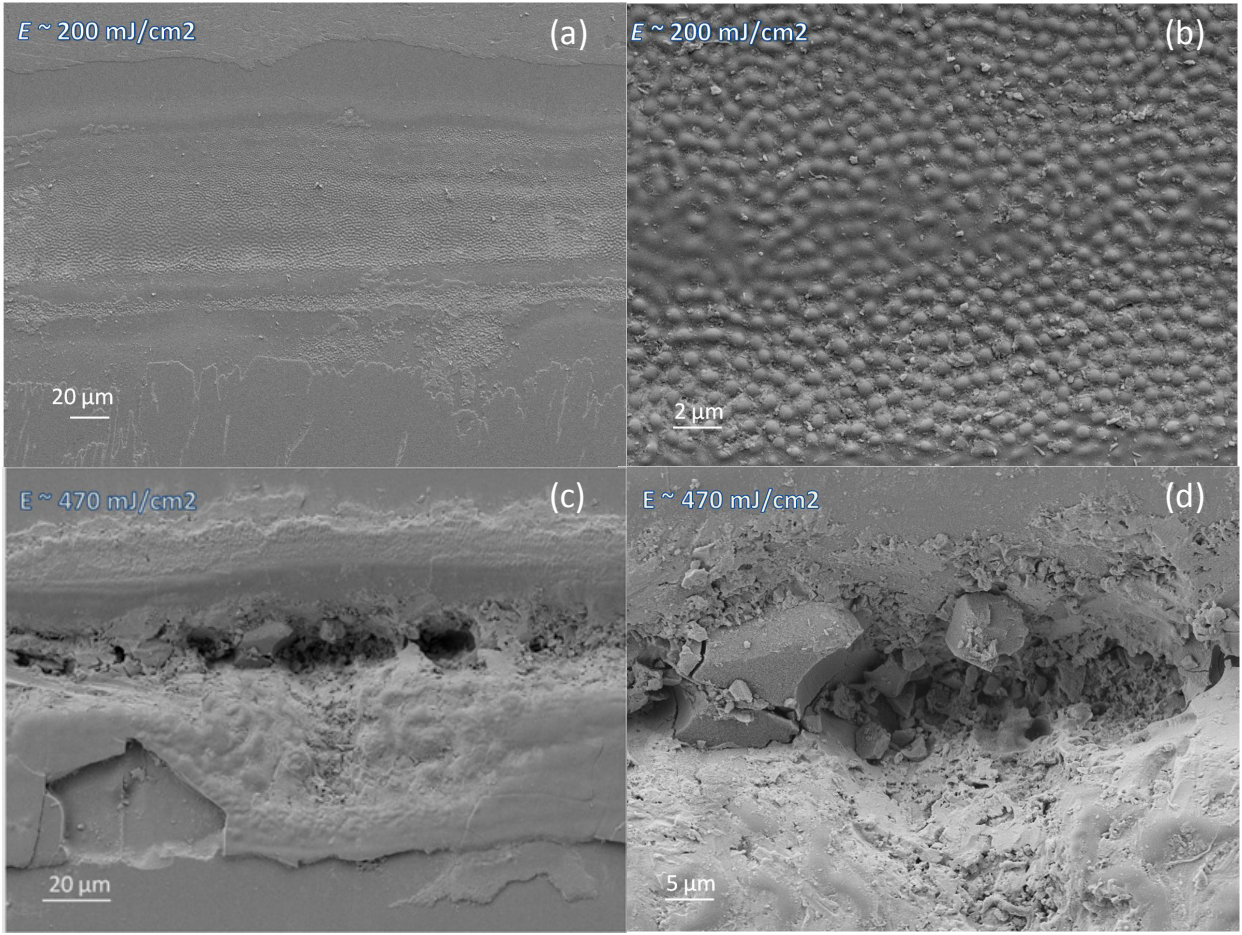

**Figure 4**. Scanning electron micrographs of the surface after fixed 193nm line exposure at two energies: (a), (b) 200m mJ/cm$^2$ and (c), (d) 470 mJ/cm$^2$.

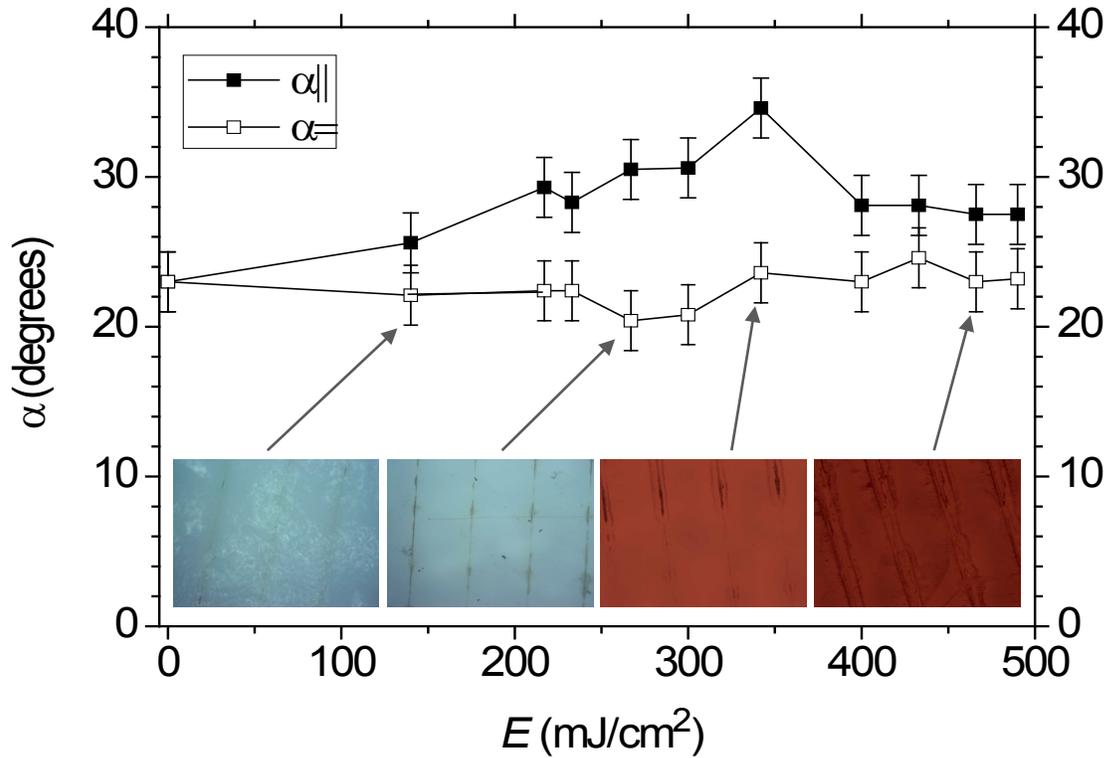

**Figure 5.** Contact angle α as a function of laser E on a set of parallel lines 0.5mm spacing; α∥ denotes α measured orthogonal to the lines, α= denotes α measured parrallel to the lines. Inset images shows optical images of a section of the surface with laser treatment at different energies – damage varies from slight coloring at low energies to surface roughening along the lines at higher energies.

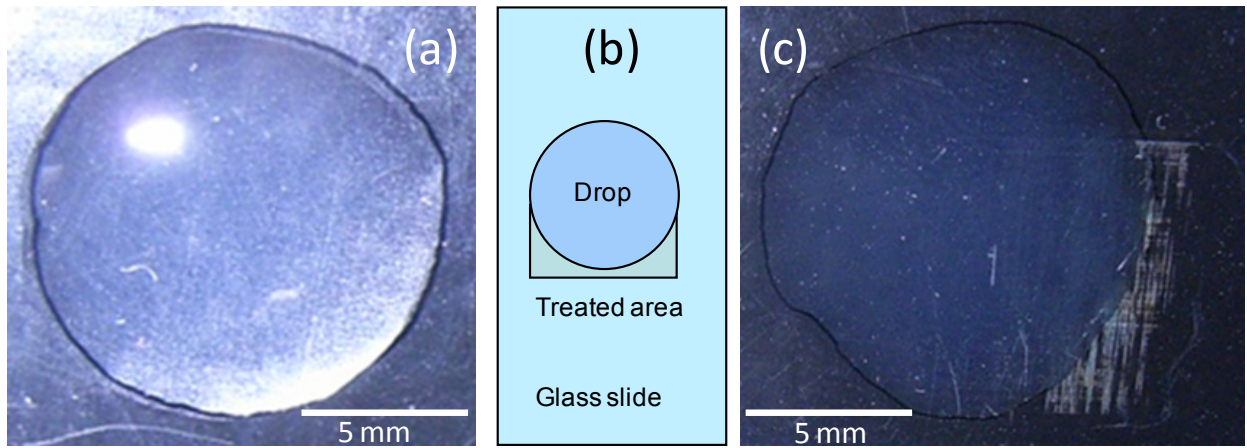

**Figure 6.** Colloidal drops containing silica nanoparticles deposited on (a) untreated surface, (b) & (c) partially over laser treated surface ($E$ = 466 mJ/cm$^2$) as depicted (b) and as actual (c). Visible, roughened damage is observed in the exposed area when the lighting is turned down.

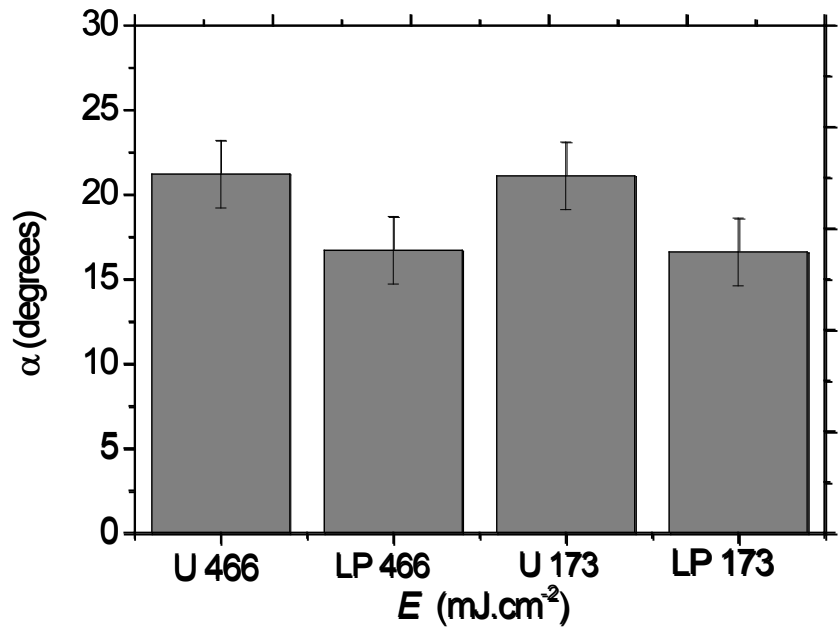

**Figure 7.** Contact angle for two samples exposed at two different energies, $E = 466$ mJ/cm$^{-2}$ & $E = 173$ mJ/cm$^2$: U – α measured on unexposed side, LP - α measured on exposed (laser processed) side.

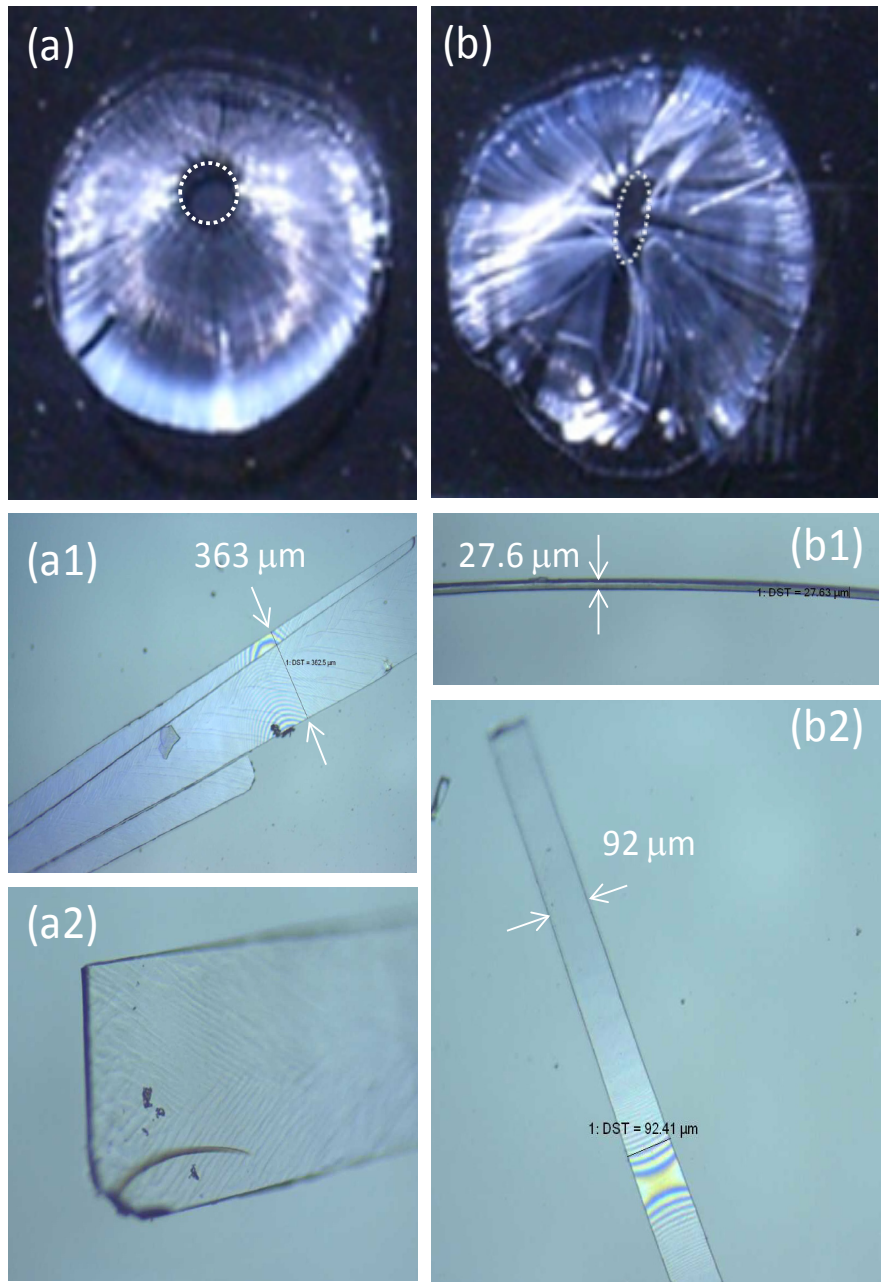

**Figure 8.** Self-assembled microwires and tapers: (a) The pristine spherical caplet produces tapered wires converging at the centre of the evaporated drop; (a1) & (a2) - dimensions are generally > 300 μm at one end and varying <100 μm at the other end; (b) The ellipsoidal caplet generated by an asymmetric contact angle leads to much more uniform wires where evaporation

has occurred in the direction of the laser processed region to centre. (b1) & (b2) show uniform wires varying between 25 and 100 μm.

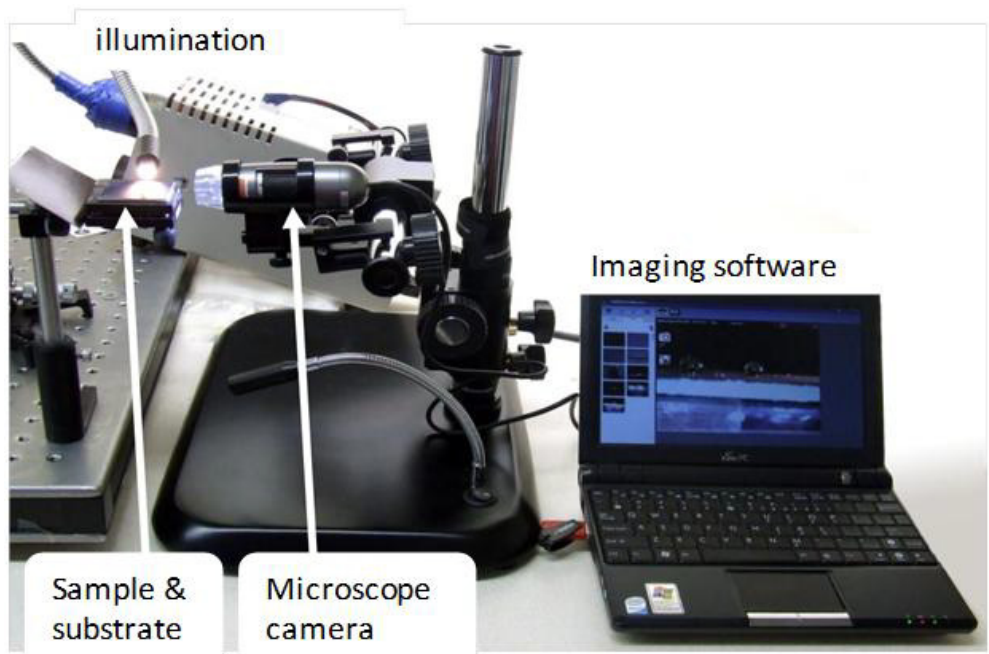

**Figure 9.** Customised contact angle goniometer used to measure contact angle.